\documentclass[journal= ancac3,manuscript=article]{achemso}

\usepackage[version=3]{mhchem}

\author{Christopher R. Iacovella$^1$}
\author{William R. French$^1$}
\author{Brandon G. Cook$^2$}
\author{Paul R.C. Kent$^3$}
\author{Peter T. Cummings$^{1,3}$}
\email{peter.cummings@vanderbilt.edu}

\affiliation[Vanderbilt University]{$^1$Department of Chemical and Biomolecular Engineering, \\$^2$ Department of Physics and Astronomy,\\ Vanderbilt University, Nashville, TN, 37235-1604 \\$^3$ Center for Nanophase Materials Sciences, Oak Ridge National Laboratory, Oak Ridge, TN 37831-6494}

\title{The Role of Polytetrahedral Structures in the Elongation and Rupture of Gold Nanowires }

\begin{document}

\begin{abstract}
We report comprehensive high-accuracy molecular dynamics simulations using the ReaxFF forcefield to explore the structural changes that occur as Au nanowires are elongated, establishing trends as a function of both temperature and nanowire diameter. Our simulations and subsequent quantitative structural analysis reveal that polytetrahedral structures (e.g. icosahedra) form within the ``amorphous'' neck regions, most prominently for systems with small diameter at high temperature. We demonstrate that the formation of polytetrahedra diminishes the conductance quantization as compared to systems without this structural motif.  We demonstrate that use of the ReaxFF forcefield, fitted to high-accuracy first principles calculations of Au, combines the accuracy of quantum calculations with the speed of semi-empirical methods.
\end{abstract}

\maketitle

The observation of conductance quantization in atomic-scale junctions 
\cite{ Muller:1992, Krans:1995, Ohnishi:1998, Yanson:1998, Guo:2011}  has fueled much interest in understanding their formation in mechanically deformed metallic nanowires (NWs) \cite{Marszalek:2000, Coura:2004,  daSilva:2004, Gall:2004, Sato:2005, Koh:2006, Pu:2007, Wang:2007, Pu:2008,  Tsutsui:2008, Lagos:2009, Tavazza:2009}, in large part due to potential applications in nano- and molecular-electronic devices.  To this extent, quantifying the structural changes that occur as a NW elongates is important for understanding how and why parameters such as initial crystal structure \cite{Coura:2004, daSilva:2004}, temperature \cite{Pu:2008}, and rate of elongation \cite{Pu:2008, Koh:2006} affect the material properties (e.g., conductance \cite{Barnett:1997, Rego:2003, Lopez:2008, Tavazza:2011}, mechanical stability \cite{Koh:2006, French:2011}, or the opening of a bandgap \cite{Nunes:2007}).  Many studies have been performed that focus on identifying changes to the crystalline structure, such as slipping and reorientation of the crystalline lattice \cite{Marszalek:2000, Wang:2007, Tavazza:2009, Pu:2008, Park:2009}, and under what conditions single-atom wide monatomic chains are likely to form \cite{Ohnishi:1998, Yanson:1998, Coura:2004, Pu:2008, Tavazza:2009}.  However, to date, there has only been limited emphasis on quantifying the formation and structure of the  ``amorphous'' or, more generally, non-crystalline domains that often appear within NWs \cite{Wang:2007, Ikeda:1999, Barnett:1997, Tavazza:2011}.   The formation of these non-crystalline structures is likely to result in significant changes to the properties of the NWs. For example, amorphous structures separating otherwise crystalline NW domains may exhibit unique dynamical properties akin to glass-forming liquids, similar to atoms at grain boundaries in bulk metals \cite{Zhang:2009}.  Moreover, non-crystalline structures may still demonstrate distinct local ordering, such as the distorted icosahedral structures predicted to form in the necks of Na NWs \cite{Barnett:1997} and the recent synthesis of bi-metallic NWs with local icosahedral ordering \cite{Jesus:2011}. This is of particular consequence as thiolated icosahedral Au NWs are predicted to behave as semiconductors under certain charge states \cite{Jiang:2009} and non-crystalline NW neck structures have recently been demonstrated to produce gradual changes in the conductance, as opposed the more typical quantization, i.e.,  ``steps'' and ``plateaus''\cite{Tavazza:2011}.

In this paper, we focus on identifying the amorphous/non-crystalline structures that form within simulated Au NWs undergoing tensile elongation.  Simulation is well suited to study this problem as it provides the full spatial and temporal coordinates of the atoms in the NW, allowing the structure to be quantified by the use of order parameters, such as those recently adapted from the field of shape matching \cite{Iacovella:2007,Keys:2011, Keys:2011b}.  Much of the previous simulation work dealing with the elongation of Au NWs has relied on either quantum mechanical \cite{Hakkinen:2000, daSilva:2004, Pu:2007, Velez:2008, Tavazza:2009, Tavazza:2010,Tavazza:2011} or, more commonly, semi-empirical methods \cite{Coura:2004, Gall:2004, Sato:2005, Wu:2006, Pu:2007, Pu:2008, Zhao:2008, Park:2009, French:2011}, requiring a compromise between the accuracy of the interatomic interactions and the accessible system size and timescale.  Quantum mechanical methods, such as density functional theory (DFT), are widely considered to provide a highly accurate description of the interactions between Au atoms, in particular, capturing the stability of 2-d planar geometries for low coordination clusters \cite{Xiao:2004, Keith:2010}.  However, the high computational cost of quantum mechanical methods often limits both the total number of independent statepoints that can be efficiently considered and the total size of the NW (typically $\sim$100 atoms or less \cite{Velez:2008, Tavazza:2009}), both of which may inadvertently bias the results.  Semi-empirical methods, such as the second-moment approximation of the tight-binding (TB-SMA) scheme \cite{Cleri:1993}, are many orders of magnitude faster than quantum mechanical methods, allowing for the efficient exploration of large numbers of statepoints and systems sizes more representative of experiment \cite{Coura:2004, Sato:2005, Pu:2007, Pu:2008}. While methods such as TB-SMA have been shown to qualitatively match the behavior of experiment in side-by-side studies \cite{Coura:2004}, semi-emperical methods are often fitted to bulk systems, and thus may produce results that are quantitatively inconsistent with theory or experiment\cite{Gall:2005}, especially for low coordination atomic-scale contacts \cite{Jarvi:2008, Keith:2010}.  In particular, the minimal energy 2-d planar structures of Au are not typically stable for semi-empirical methods \cite{Keith:2010} which may incorrectly bias NW simulations towards forming 3-d isomers.   

In order to bridge the gap between semi-empirical and quantum mechanical methods, we use the ReaxFF reactive forcefield \cite{goddard} which is designed to provide accuracy approaching quantum mechanical methods, but at a substantially reduced computational cost.  ReaxFF forcefields are derived by fitting analytical functions to DFT bonding curves\cite{goddard, Keith:2010}.  The parameters derived for Au by Keith, \textit{et al.}\cite{Keith:2010} have been demonstrated to satisfactorily capture the expected behavior of both bulk and low coordination states \cite{Keith:2010}.  In particular, this parameterization has been shown to closely reproduce the energy and stability of low and minimal energy clusters of Au\cite{Keith:2010}.  In this work, we show that ReaxFF is a substantial improvement over TB-SMA for the simulation of NW structures by comparing the absolute and relative energy scaling to DFT calculations.  We perform comprehensive molecular dynamics (MD) simulations\cite{Plimpton:1995} with ReaxFF to investigate the behavior and properties of [100] orientated FCC Au NWs undergoing tensile elongation in vacuum, which we also compare to simulations performed using TB-SMA.  We report, for the first time, the formation of well-ordered polytetrahedral local structures (e.g., full and partially coordinated icosahedra \cite{Frank:1958}) within the non-crystalline domains of the NWs.  We quantify the formation of these structures by using the R$_{ylm}$ shape matching method \cite{Iacovella:2007, Iacovella:2008, Keys:2011, Keys:2011b} based on spherical harmonics\cite{Steinhardt:1983}, which we use to construct trends as a function of temperature and NW diameter.
We further calculate the zero-bias conductance of our simulation trajectories to demonstrate the impact that the formation of polytetrahedra has on the conductance behavior. 

\section{Results}

\subsection{Validation of ReaxFF for Au NWs}

We first assess the energetic description of Au NWs predicted by ReaxFF by comparing to DFT and TB-SMA.  We report the energy for [100] oriented periodic FCC NWs with diameters, D=1.1, 1.5, and 1.9 nm, as well as an elongation sequence of a periodic 0.85 nm NW.  The elongation sequence is representative of the low coordination structures typically found in the necks of mechanically deformed NWs.  We perform DFT calculations that make use the generalized gradient approximation by Perdew, Burke, Ernzerhof (PBE-GGA) and additionally include spin-orbital coupling (see Methods). All reported DFT energy values are calculated using structures that have been fully and self-consistently energy-minimized.  

\begin{figure}[h] %  figure placement: here, top, bottom, or page
   \centering
   \includegraphics[width=5.0in]{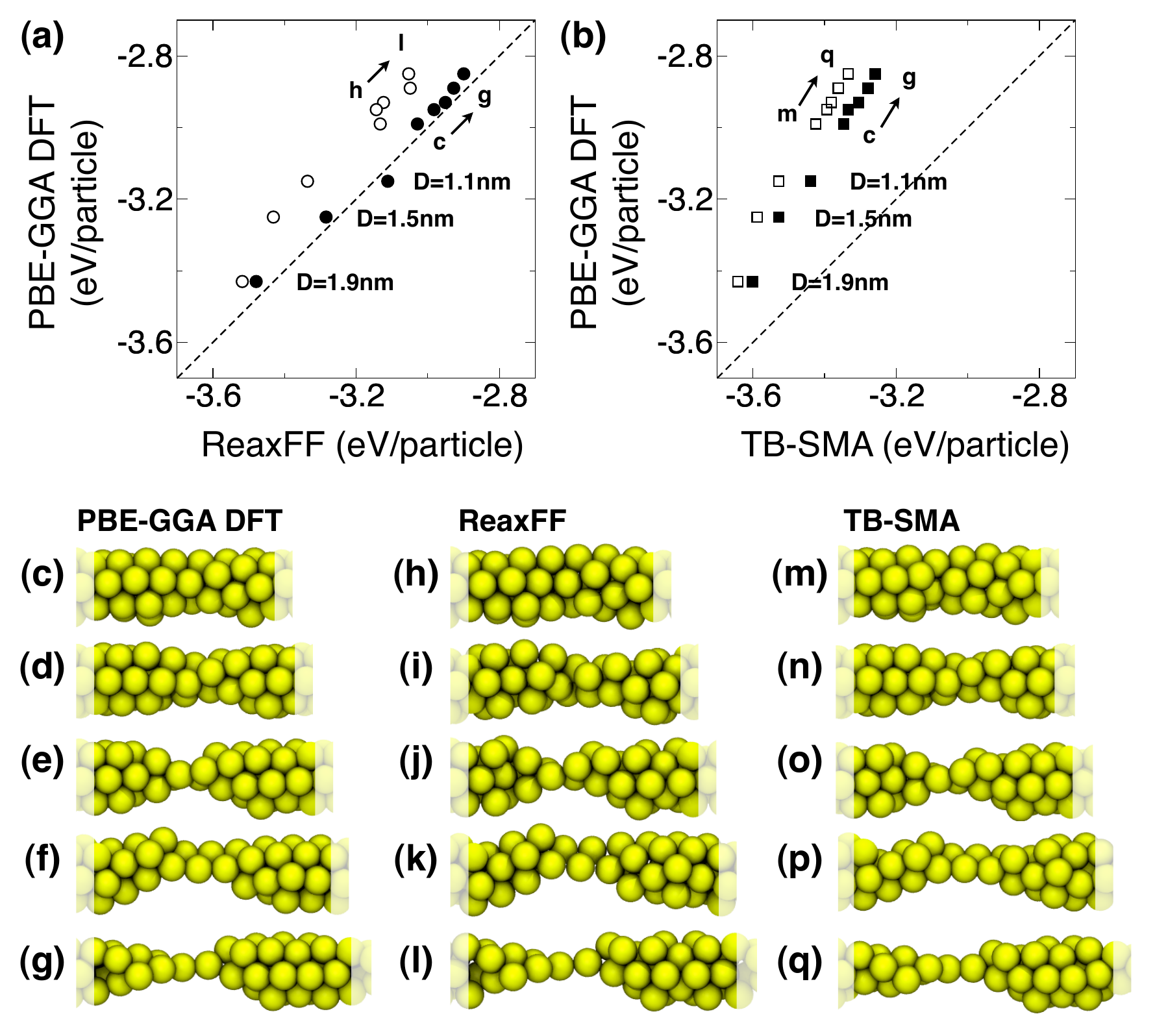} 
   \caption{\textbf{(a)} DFT energy vs. ReaxFF energy.  \textbf{(b)} DFT energy vs. TB-SMA energy.  In both (a-b) filled symbols correspond to the energy calculated of structures that have been energy-minimized with DFT, open symbols correspond to energy of structures energy-minimized with each of the methods (e.g., DFT energy-minimized and ReaxFF energy-minimized).  Snapshots of the energy-minimized configurations of the elongation sequence for: \textbf{(c-g)} DFT, \textbf{(h-l)} ReaxFF, and \textbf{(m-q)} TB-SMA.}
   \label{fig:energy}
\end{figure} 

In \ref{fig:energy}a we plot the energy calculated with DFT vs. the energy calculated with ReaxFF.  In this plotting scheme, an ideal match between DFT and ReaxFF occurs if the data point falls along the line $y = x$.  In \ref{fig:energy}a, filled symbols represent the energy calculated using the coordinates energy-minimized with DFT (i.e., when calculating the ReaxFF energy, atoms are static and arranged according to the configuration generated with DFT). The DFT energy-minimized configurations for the elongation sequence are shown in Figures \plainref{fig:energy}c-g.  We observe that ReaxFF is capable of accurately reproducing the per particle energy predicted by DFT, where the magnitude of the difference is less than 0.05 eV per particle and all data points fall along the line $y = x$.  In \ref{fig:energy}a, data points with open symbols correspond to energy-minimized structures predicted by the two different treatments (i.e., the DFT energy corresponds to the structure energy-minimized with DFT and the ReaxFF energy corresponds to the structure energy-minimized with ReaxFF).  The ReaxFF structures are generated starting from the DFT optimized structures.  The energy-minimized elongation sequence for ReaxFF is shown in Figures \plainref{fig:energy}h-l, where we observe minor changes as compared to the DFT structures in Figures \plainref{fig:energy}c-g.  Note, the preferred interatomic spacings will vary slightly between methods, thus we use a two stage energy minimization for ReaxFF, first allowing the box volume to change to minimize pressure (which scales atom positions), then allowing atom positions to change in order to minimize the potential energy.   For minimized configurations, ReaxFF predicts a lower energy structure than DFT for all statepoints. However the values are still close, where the magnitude of the deviation varies from 0.09 to 0.20 eV/particle.  %Note, previous PBE-GGA DFT calculations of bulk Au have been reported to deviate by 0.08 eV/particle when compared to experiment.  Specifically, the energy measured with DFT is given as -3.85 eV/particle\cite{Keith:2010} and experiment measured to be -3.93 eV/particle\cite{Smith:1976}.

\ref{fig:energy}b is constructed in the same manner as \ref{fig:energy}a, but using TB-SMA rather than ReaxFF.  Again, filled symbols represent the energy calculated using the DFT energy-minimized structures.  TB-SMA fails to closely reproduce the energy, predicting lower values than DFT for all structures.  Here the magnitude of the differences ranges from 0.17 to 0.41 eV per particle. Open symbols again represent the energy of structures minimized with the two methods.  We use a similar procedure for TB-SMA as we did for ReaxFF. The energy minimized structures of TB-SMA are shown in \ref{fig:energy}m-q, where we observe only minor differences when compared to either DFT or ReaxFF structures.  The most noticeable difference is observed in \ref{fig:energy}q, where the left hand side of the NW compresses into a more locally dense state.  For energy-minimized structures, the energy predicted by TB-SMA is shifted to even more negative values, where now the magnitude of the energy difference with DFT ranges from 0.21 to 0.47 eV per particle.

In addition to comparing the absolute energy, we examine the relative change in energy ranging from the highest coordination state (D=1.9nm NW) to the lowest coordination state (i.e., the final elongated structure, shown in Figures \plainref{fig:energy}g, l and q).  This can be assessed by examining the slope of the data points shown in Figures \plainref{fig:energy}a and b; a slope of unity provides an ideal energy change with respect to DFT.  For the ReaxFF calculations in \ref{fig:energy}a, we observe a slope 0.99 for the static structure (filled symbols) and 1.11 for the energy-minimized structures (open symbols), using linear regression.  Thus, even though the absolute energies are slightly different for the energy-minimized states (open symbols), the relative change predicted by ReaxFF is in very good agreement with DFT.  For the TB-SMA calculations in \ref{fig:energy}b, we observe a slope of 1.63 for static calculations (filled symbols) and a slope of 1.81 for energy-minimized structures (open symbols).  The strong deviation from unity indicates that TB-SMA fails to properly predict the energy loss associated with a reduction in average coordination number, which may strongly influence the structural evolution of the NW.  In previous work comparing common semi-empirical methods for the elongation of finite Au NWs, TB-SMA was demonstrated to provide the best agreement with DFT in terms of the relative energy change\cite{Pu:2007}.  Thus it appears that ReaxFF is an improvement over TB-SMA, providing a better description of both the relative energy change and absolute energy for Au NWs.

\subsection{The formation of polytetrahedral structures}
\begin{figure}[h] %  figure placement: here, top, bottom, or page
   \centering
   \includegraphics[width=3.3in]{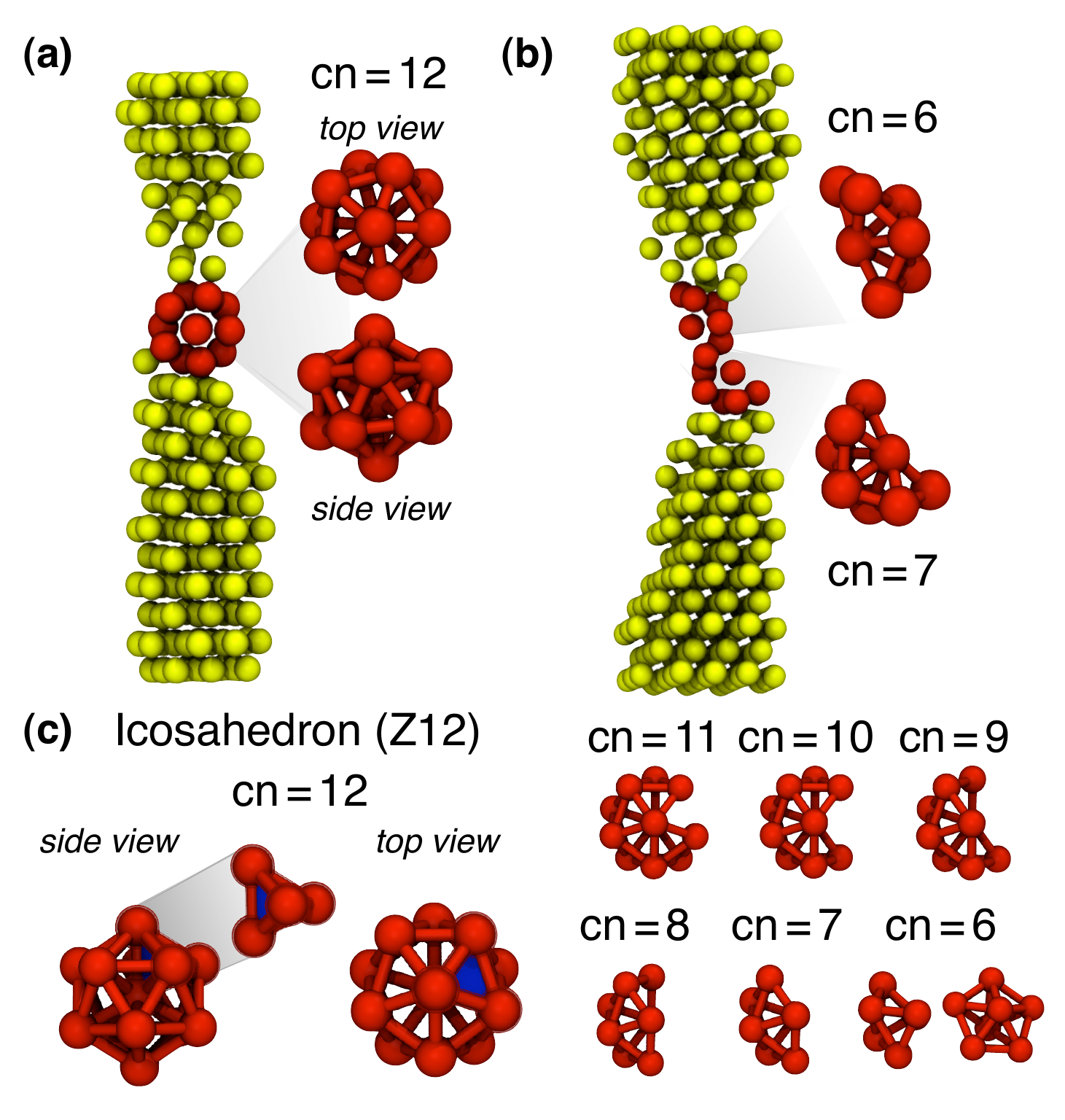} 
   \caption{%\textbf{(b)} Rendering of a Z14 cluster and partial clusters with cs=15 (perfect) to cs=10. 
Representative ReaxFF simulation snapshots of NWs at T=298K elongated at 0.1 m/s for  \textbf{(a)}  D=1.1nm NW elongated by 12.2 \AA~ and \textbf{(b)} D=1.5nm NW elongated by 21.9 \AA~.  In a and b, atoms are rendered at $\sim$ 75$\%$ of their true size with individual polytetrahedral structures highlighted within the NW and shown at the right. 
\textbf{(c)} Rendering of a Z12 structure, i.e., an icosahedron, with an individual tetrahedral subunit highlighted. Partial Z12 clusters are shown as a function of coordination number, cn, varying from fully coordinated, cn=12, to cn=6. Note, cn=6 shows two different partial clusters commonly observed.   }
   \label{fig:polyhedra}
\end{figure}

We perform simulations of cylindrical Au NWs of length 3.6 nm, with diameters, D=1.1, 1.5 and 1.9 nm (177, 317, and 589 atoms, respectively) for a variety of statepoints, calculating 10 independent simulations for each parameter combination (see Methods for more details).  Here we focus on ReaxFF-based simulations but also include TB-SMA-based simulations for comparison. In these ReaxFF simulations we observe that, as a typical Au NW elongates, a small region begins to ``neck,'' becoming narrower than the original diameter.  The neck forms in response to the applied tensile load, localizing many of the structural changes to this neck region, allowing the bulk of the NW to relax its configuration.  Within the necks of our simulated NWs, we often observe the formation of non-crystalline structures with distinct ring- and crescent-like geometries.  \ref{fig:polyhedra}a shows a representative ReaxFF simulation snapshot of a D=1.1 nm NW at T=298K,  shortly after neck formation; an individual ring-like structure is highlighted in the snapshot with an enlarged version, extracted from the system, shown to the right.  \ref{fig:polyhedra}b shows a representative snapshot of a D=1.5nm NW where two crescent-like structures are highlighted and extracted. These ring- and crescent-like structures are polytetrahedral in nature, where the local structure of the atoms can be decomposed solely into individual tetrahedral subunits\cite{Frank:1958}. Specifically, the highlighted structure in \ref{fig:polyhedra}a is best classified as an icosahedron. An icosahedron is constructed of 13 particles arranged as two splayed pentagonal pyramids (composed of 20 total tetrahedral subunits), demonstrating 5-fold symmetry\cite{Frank:1958}; an ideal icosahedron is shown in \ref{fig:polyhedra}c. The two highlighted clusters in \ref{fig:polyhedra}b are partially coordinated icosahedral clusters, that maintain the same basic geometry of the full cluster, but with $X$ number of particles removed (i.e., several tetrahedral subunits removed)\cite{Iacovella:2007}. Ideal partial clusters of various coordination are shown in \ref{fig:polyhedra}c.  TB-SMA simulations also demonstrate polytetrahedral structures within the necks of the elongated NWs, similar to Figures \plainref{fig:polyhedra}a and b.   

\begin{figure}[h] %  figure placement: here, top, bottom, or page
   \centering
   \includegraphics[width=3.4in]{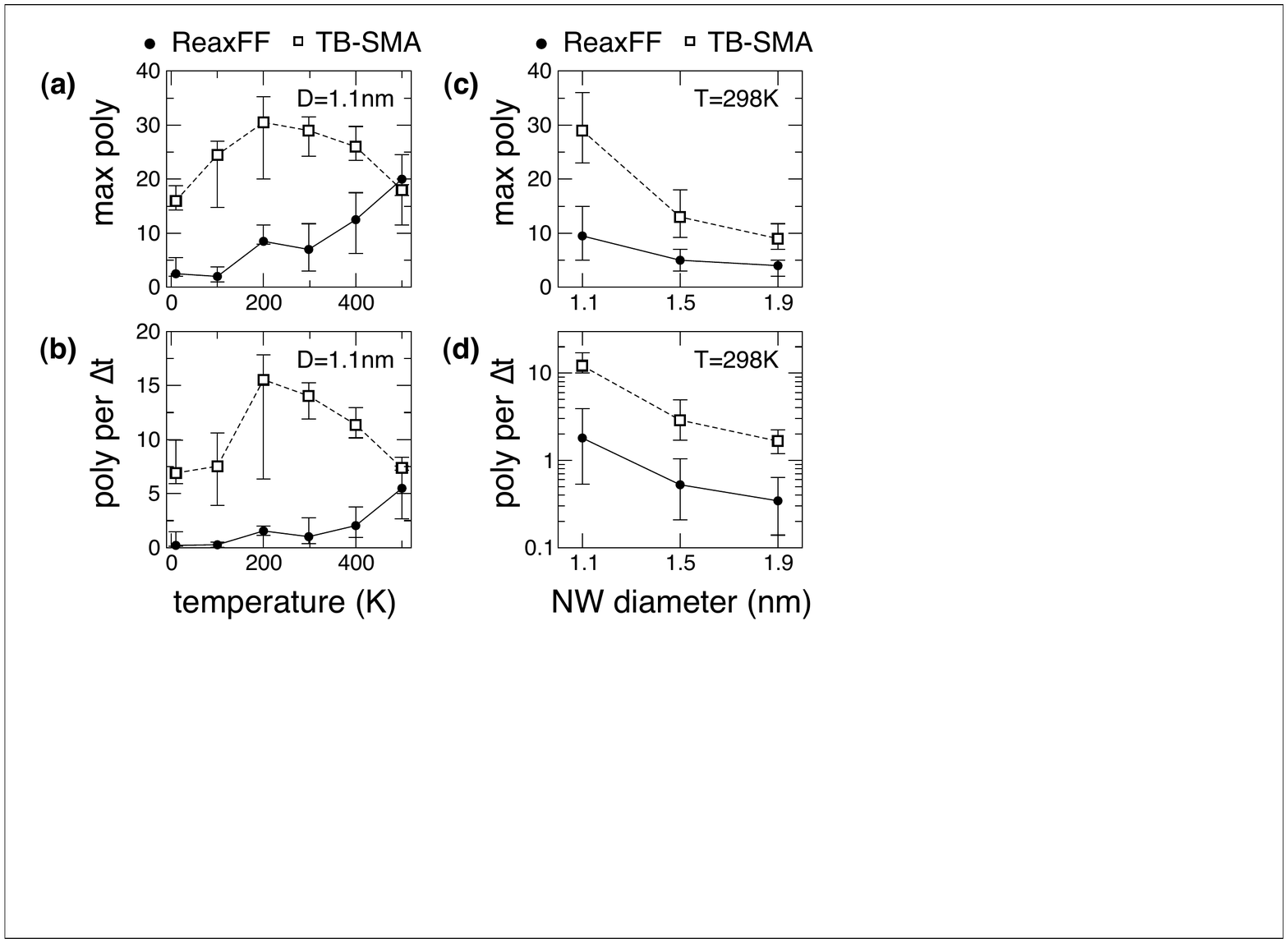} 
   \caption{Quantification of polytetrahedral structures.   \textbf{(a)} The maximum number of polytetrahedral clusters observed concurrently in a trajectory and \textbf{(b)} the number of polytetrahedra per timestep, both for D=1.1nm and shown as a function of temperature for NWs pulled at 1 m/s, calculated from 120 total independent simulations.  \textbf{(c)} The maximum number of polytetrahedral clusters observed concurrently in a trajectory and \textbf{(d)} the number of polytetrahedra per timestep, both at T=298K and shown as a function of diameter, calculated from 400 total independent simulations.  Data for (c-d) is presented as the aggregate of simulations conducted from 0.1 to 5 m/s, as a strong rate dependence was not observed. All data points correspond to the median and error bars correspond to 1st and 3rd quartiles. The legends shown at the top are applicable to all plots.}
   \label{fig:poly_fd}
\end{figure} 

To quantify the structures formed in our simulations and establish trends, we use the $R_{ylm}$ method to determine which atoms are in polytetrahedral local environments (see Methods). We use this quantitative information to construct three metrics: (1) the maximum number polytetrahedra that occur concurrently during the course of a simulation trajectory, (2) the number of polytetrahedra per timestep in a simulation (calculated by normalizing the total number of polytetrahedra observed by the total number of statepoints considered), and (3) the coordination number (cn) histogram of polytetrahedral atoms.

We first report the behavior of thin NWs with D=1.1 nm elongated at a rate of 1 m/s for T=\{10, 100, 200, 298, 400, 500\} K. Figures \plainref{fig:poly_fd}a and b plot the maximum number of polytetrahedra and the polytetrahedra per timestep, respectively. In all cases, we report median values, where error bars correspond to the first and third quartiles, since the data is not necessarily well described by a Gaussian distribution.  ReaxFF simulations demonstrate a steady increase in both metrics as temperature is increased, i.e., both the maximum size of the polytetrahedral domains and the likelihood of finding polytetrahedra increases as T increases.  The polytetrahedral motif is not prominent for ReaxFF systems at low temperature, i.e., T $<$ 200K.  While the behavior of the ReaxFF and TB-SMA potentials converge for high temperature, TB-SMA demonstrates markedly different trends and behavior.  Overall, we observe that TB-SMA predicts a significantly larger value of both the maximum number of polytetrahedra and polytetrahedra per timestep than the ReaxFF simulations. TB-SMA predicts two regimes with an approximate crossover of T=200K. TB-SMA systems at T=10 and 100K tend to maintain a predominantly FCC structure, whereas higher temperature states transition into A3 core-shell helical structures \cite{Wang:2001} upon moderate amounts of elongation.  For example, at T=298K, 70$\%$ of the TB-SMA simulations transform into predominantly A3 structured NWs by the time the NW has been elongated by 5\AA.  The A3 structure is locally composed of tetrahedral subunits that closely match the polytetrahedral structures in our $R_{ylm}$ reference library.   However, we note that even for T$<$200K, where A3 structures are not found with either potential,  ReaxFF still forms fewer polytetrahedra than TB-SMA, by a factor of $\sim$10 for both metrics.  The predominance of crystalline structures at low temperature predicted by ReaxFF is more consistent with results of zero temperature DFT calculations performed for the elongation of [110] and [111] Au NWs \cite{Tavazza:2009, Tavazza:2011}.  

Figures \plainref{fig:poly_fd}c and d plot the median value of the maximum number of polytetrahedra and polytetrahedra per timestep, respectively, for NWs as a function of NW diameter; these simulations are performed at T=298K, as this statepoint demonstrates appreciable numbers of polytetrahedra for both potentials and temperatures in this vicinity are common in experiment \cite{Marszalek:2000, Coura:2004}.  The values reported in Figures. \plainref{fig:poly_fd}c and d correspond to the aggregate of simulations performed for rates 0.1 to 5 m/s, as we did not observe strong variability with rate of elongation. We observe that the median value of both metrics decreases with increasing diameter for both ReaxFF and TB-SMA.  That is, the likelihood of forming polytetrahedra decreases with increasing NW diameter.  In all cases, ReaxFF-based simulations predict smaller median values for the two metrics than TB-SMA, with the largest differences occurring for the D=1.1nm NWs.  The differences we observe between ReaxFF and TB-SMA are likely related to the energy scaling we previously explored, where TB-SMA overpredicts both the absolute magnitude and relative energy of lower-coordination states.

\begin{figure}[h] %  figure placement: here, top, bottom, or page
   \centering
   \includegraphics[width=2.0in]{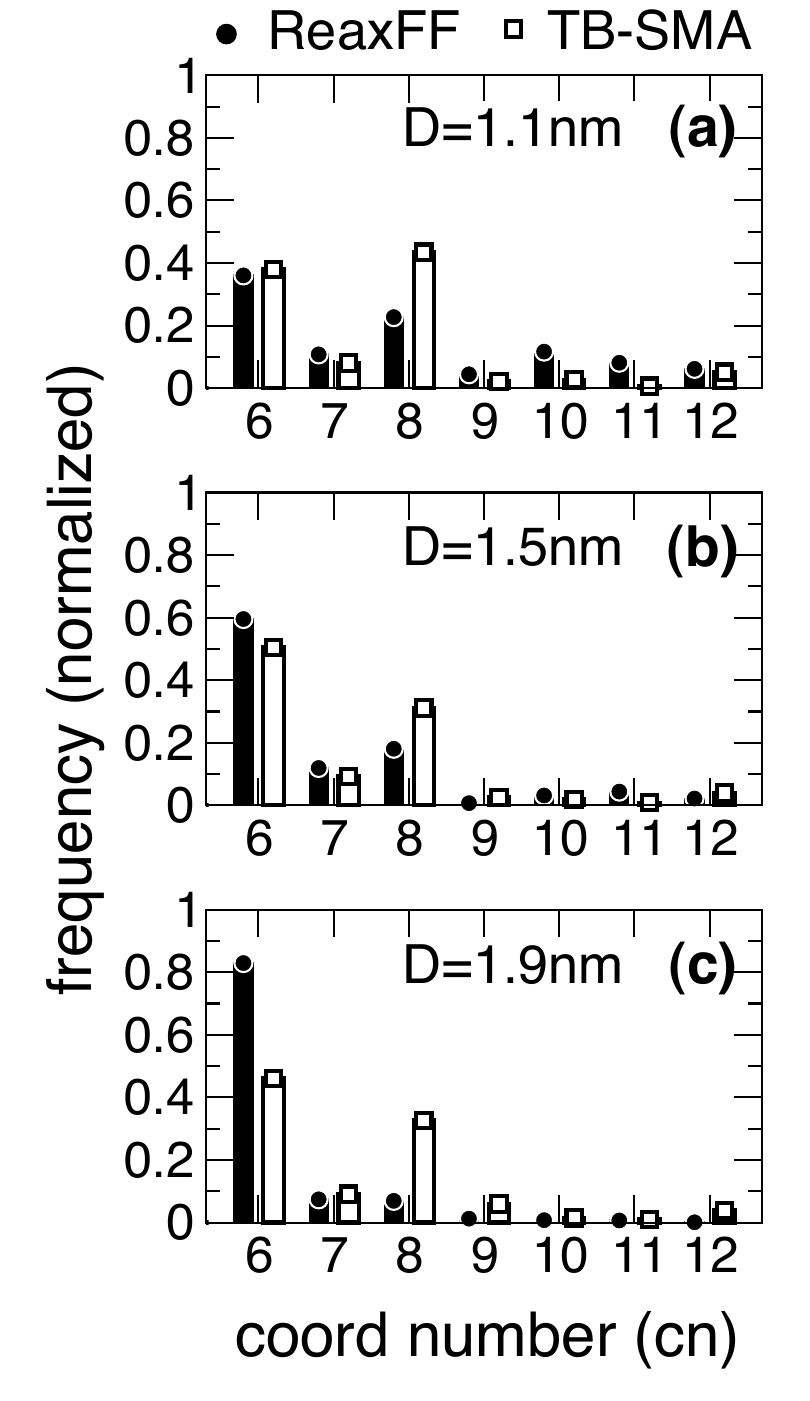} 
   \caption{Histograms of the coordination number (cn) of polytetrahedral structures for NWs with \textbf{(a)} D=1.1nm, \textbf{(b)} D=1.5nm, and \textbf{(c)} D=1.9nm. All data is at T=298K and is presented as the aggregate of simulations conducted from 0.1 to 5 m/s.  The legend at top is applicable to all plots.}
   \label{fig:CN}
\end{figure} 

\ref{fig:CN} plots histograms of the coordination number of atoms identified as having polytetrahedral local ordering for NWs at T=298K (corresponding to the data presented in Figures \plainref{fig:poly_fd}c and d). Histograms of ReaxFF simulations demonstrate a diameter dependence; D=1.1nm NWs show two key peaks at cn=6 and 8, but the relative fraction of cn=6 strongly increases with increasing diameter. For D=1.9 nm NWs simulated with ReaxFF, $\sim$ 80 \% of the observed polytetrahedral clusters have cn=6. Histograms generated from TB-SMA simulations demonstrate two clear peaks at cn=6 and 8, encompassing $\sim$ 80 \% of the total polytetrahedral structures, with much less dependence on diameter.  %As we discuss later, our DFT calculations suggest that partial icosahedra with cn=6 and 8 are locally stable

\subsection{Zero-bias conductance}
We calculate the zero-bias conductance of several representative trajectories generated with ReaxFF and TB-SMA in order to assess the impact of polytetrahedra, i.e., is there a characteristic conductance ``fingerprint'' associated with their formation.  As described in the Methods section, we use a combined DFT-non-equilibrium Green's function technique to calculate the conductance\cite{PhysRevB.81.115412}, which we report in terms of $G_{0} = 2e^2/h$, where $e$ is the charge of an electron and $h$ is Planck's constant.  In experiment, a typical conductance trace demonstrates a ``staircase'' behavior, with abrupt plateaus and drops as a function of elongation. The abrupt changes in the conductance coincide with structural rearrangements of the atoms, such as thinning of the NW neck.   For very small cross-sections, the steps often occur at roughly integer multiples ($m$) of $G_{0}$. It is important to note that histograms of conductance tend to demonstrate a substantial spread around each of these integer peaks \cite{costakramer:1997} and, as a result, non-integer values are also regularly observed in experiment.  Recent work has correlated the formation of non-integer plateaus to the difference in dimensionality of the local structures within the NW neck (i.e., the difference between 3-, 2-, and 1-dimensional atomic arrangements)\cite{Tavazza:2011}. 

\begin{figure}[h] %  figure placement: here, top, bottom, or page
   \centering
   \includegraphics[width=4.5in]{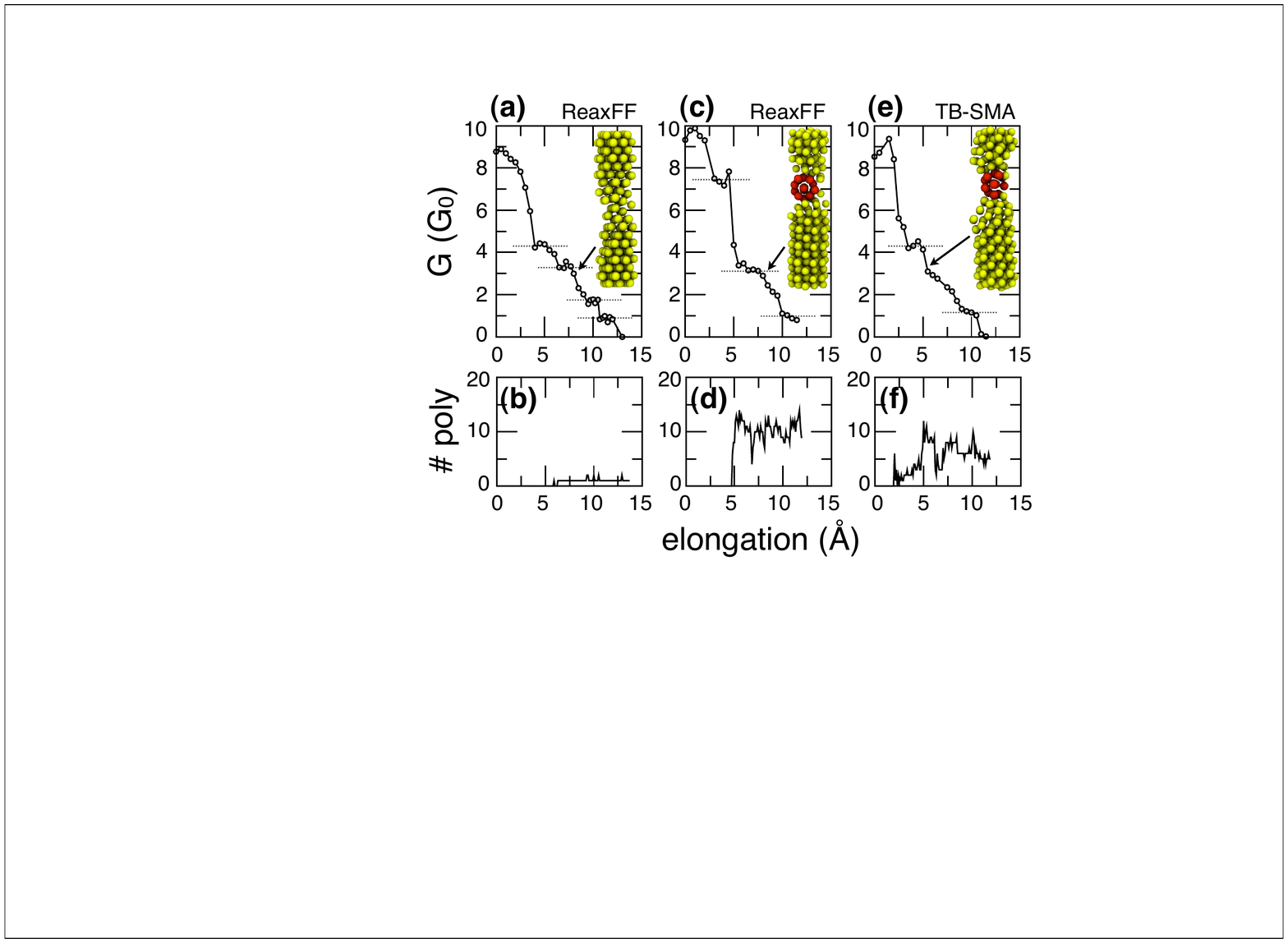} 
   \caption{\textbf{(a)} Conductance and \textbf{(b)} number of polytetrahedra as a function of elongation of a ReaxFF NW at T=298K. \textbf{(c)} Conductance and \textbf{(d)} number of polytetrahedra of a ReaxFF NW at T=10K.   \textbf{(e)} Conductance and \textbf{(f)} number of polytetrahedra of a TB-SMA NW at T=10K. All data corresponds to NWs with D=1.1nm elongated at 1 m/s. Plateaus are highlighted with horizontal dotted lines. Simulation snapshots for configurations with G$\sim$3$G_0$ are inset in each figure, where c and e each highlight a polytetrahedral structure in the neck (red).}
   \label{fig:conductance}
\end{figure}

\ref{fig:conductance} plots the conductance and number of polytetrahedra as a function of elongation for trajectories of D=1.1nm NWs elongated at 1 m/s, generated with both ReaxFF and TB-SMA.  We first consider a ReaxFF trajectory at T=298K that does not demonstrate any significant polytetrahedral ordering, as shown in Figures \plainref{fig:conductance}a-b.  We observe distinct step-wise behavior with conductance plateaus at roughly integer multiples of $G_0$ ($m \sim$ 1,2, 3 and 4 G$_0$, highlighted in \ref{fig:conductance}a with horizontal lines).  Additionally, the length of these plateaus range from 1.5-3.5 \AA~ which matches the approximate plateau length of 1-4\AA~ seen in experiments at T=4K \cite{Yanson:1998} and T=298K \cite{Marszalek:2000}.  A snapshot of the NW at $\sim$3$G_0$ is inset in \ref{fig:conductance}a showing a crystalline neck structure. 

Figures \plainref{fig:conductance}c-d plot data for a NW trajectory generated with ReaxFF at T=10K.  As before, we observe step wise behavior in the conductance.  In contrast to the previous trajectory, we now observe the formation of an appreciable number of polytetrahedral structures, as shown in \ref{fig:conductance}d. The formation of polytetrahedra appears to have a distinct impact on the conductance.  For example, while we observe plateaus at $m \sim$3 and 1 $G_0$ in \ref{fig:conductance}c, there is a gradual change in conductance between these two steps rather than a plateau at 2 $G_0$.  The onset of the formation of polytetrahedral structures at $\sim$5\AA~correlates with the start of the conductance plateau at 3$G_0$. Visually, we observe that at the start of the gradual change (i.e, 3$G_0$), the narrowest region of the neck is spanned by a polytetrahedral structure $\sim$ 3 atoms wide; a shapshot is inset in \ref{fig:conductance}c.  

We further assess the impact of polytetrahedra by examining a trajectory at T=10K generated with TB-SMA, as shown in Figures \plainref{fig:conductance}e-f.  Recall that TB-SMA over-predicts the tendency to form polytetrahedra as compared to ReaxFF, making it ideal for assessing their impact.  We still observe step-like behavior, but now only see plateaus at $m \sim$ 4 and 1 $G_0$.  Again, we observe a rather gradual transition from 3$G_0$ to 1$G_0$.  While the onset of polytetrahedra appears at ~2.5 \AA~ of elongation, the most significant peak occurs at $\sim$5\AA~ (see \ref{fig:conductance}f), which correlates to the start of the gradual change in conductance. As was noted in \ref{fig:conductance}c, we also observe that at $\sim$3$G_0$ (i.e., the start of the gradual change), the narrowest region of the neck is spanned by a polytetrahedral structure $\sim$ 3 atom diameters across (a shapshot is inset in \ref{fig:conductance}c).   Thus it appears that the gradual conductance changes we observe are associated with the formation of polytetrahedra that span the cross-section of the neck.  This is likely related to different modes of mechanical deformation exhibited by tetrahedral structures verses crystals; the mechanical behavior of related core-shell NWs as compared to crystals has previously been explored in Ref. \cite{Gall:2004}. 

Our results regarding the impact of polytetrahedral neck structures on the conductance is supported by recent simulations of Tavazza, \textit{et al.} that used DFT to investigate the elongation of [110] and [111] Au NWs at zero temperature \cite{Tavazza:2011}.  In that work, nearly identical conductance behavior was observed for systems that demonstrated non-crystalline neck structures, i.e., gradual changes from 3$G_0$ to 1$G_0$ \cite{Tavazza:2011}.  Similar conductance traces were also observed in experiments at T=4K \cite{Tavazza:2011}.  Additionally, it was noted that systems exhibiting this gradual conductance change occurred in the minority of simulations and experiments, suggesting that non-crystalline structures are rare at these very low temperatures\cite{Tavazza:2011}.  As previously shown in Figures \plainref{fig:polyhedra}a and b, our ReaxFF simulations also predict that the non-crystalline polytetrahedral structures are a rare structural motif at low temperature.

\section{Discussion}

The formation of polytetrahedral structures within the necks of the elongated Au NWs may be surprising as icosahedra are not predicted to be the lowest energy structures \cite{Li:2007,Gruber:2008, Pundlik:2011}.  Generally speaking, planar structures are predicted to be the minimal energy states for small, isolated clusters of Au \cite{Xiao:2004,Li:2007}. However, this does not imply that icosahedra, or other low, but not minimal energy structures are inherently unstable.  To assess the stability of the polytetrahedral structures, we perform conjugate gradient energy minimization using DFT of the isolated icosahedral clusters shown in \ref{fig:polyhedra}c.  Starting from both ideal structures and those generated with ReaxFF, we find that clusters with cn=6, 8, 9, and 12 maintain their original geometry at the end of the minimization, suggesting that these structures are locally stable.  We also note that systems that start as partial icosahedra with cn=10 and 11 relax their structures into more ``flattened'' states that, in 3 out of 4 calculations, contain a distinct polytetrahedral pentagonal di-pyramind (i.e., a cn=6 polytetrahedral structure). Recall that the cn=6 structure is the most common polytetrahedral configuration found in our simulations.  Furthermore, polytetrahedral structures of Au are well reported in experiment and theory, including monolayer protected icosahedral clusters with cn=12\cite{Menard:2006}, phosphone-thiolate-protected partial-polytetrahedral Au structures\cite{Walter:2008}, and bi-icosahedral clusters of both face-sharing \cite{Pei:2008} and vertex-sharing icosahedra \cite{Shichibu:2007}.  

It is important to note that the polytetrahedral structures we observe in our simulations are not isolated clusters, but rather occur within the context of a larger, mechanically deformed NW.  The average coordination number of atoms in an isolated cluster will be lower than an equivalent configuration that occurs within a NW.  Additionally, we must also consider that these structures form as a result of mechanically deforming a NW via an applied tensile load.  Previous experiments of NWs pulled at a constant rate demonstrate a saw-tooth like behavior in the mechanical force on the NW as a function of elongation \cite{Agrait:1995} and thus the local forces of the atoms may be far from equilibrium and constantly changing.  The original structure of the NW will also dictate how it can respond to tensile loads \cite{Gall:2004, Coura:2004}. For example, it has been demonstrated that the likelihood of forming monatomic Au chains in elongated Au NWs is influenced by the original orientation of the crystalline lattice (i.e., [111] vs. [110] vs. [100]) \cite{Coura:2004}.  Focusing only on minimal potential energy structural motifs also ignores the importance of thermal effects.  Recall that in Figures \plainref{fig:poly_fd}a-b we observed that the formation of polytetrahedral structures is very unlikely at low temperature, but increases with temperature.  Previous work from our group proposed a universal energy release mechanism that outlines the strong role that temperature can play in determining how a NW responds to the applied tensile load \cite{Pu:2008}.  

Additionally, Au is ductile, which manifests itself as a tendency for the NW to remain in a cylindrical, connected structure by thinning within the neck region.   As the neck of the NW thins, polytetrahedral structures may become the optimal configurations, since they are able to span narrow, cylindrical-like geometries by forming face-sharing polytetrahedral helices \cite{Zheng:1990, Lidin:1996, Jiang:2009} or vertex-sharing structures \cite{Shichibu:2007, Jiang:2009}.  This was demonstrated for particles with van der Waals interactions, where a transition was observed from FCC/HCP structures to polytetrahedral structures upon decreasing the diameter of the confining cylinder \cite{Iacovella:2007}. Similar behavior was also demonstrated for Au, where tetrahedral helices were predicted to be the optimal structure for small diameter NWs whereas FCC structures were optimal for large diameter NWs \cite{Wang:2001}.  Recent experimental work has demonstrated the synthesis of thin Au/Ag NWs that are locally constructed of face-sharing icosahedra\cite{Jesus:2011}, additionally demonstrating this packing motif for metallic NWs.  Furthermore, the formation of polytetrahedral structures within the necks of elongating NWs is not without precedent. Previous DFT calculations of the mechanical deformation of [100] oriented BCC Na NWs demonstrated the formation of distorted full and partial icosahedra within the NW neck \cite{Barnett:1997}.  This may suggest that the formation of polytetrahedra is a more universal relaxation mechanism for mechanically deformed NWs.  

The formation of polytetrahedral structures may have important implications regarding the electronic properties of the NW.  As we saw in \ref{fig:conductance}, the formation of polytetrahedra diminishes the quantization of the conductance traces.  Combined with the increased likelihood of forming polytetrahedra as temperature increases, this may be an important to consider when engineering nanoelectronic devices designed to operate at, e.g., room temperature.  The impact of polytetrahedra may be even more substantial.  The work of Jiang, \textit{et al.} predicts that thiolated vertex-sharing icosahedral Au NWs behave as a semiconductor for the -2 and +2 charge states\cite{Jiang:2009}.  This suggest that Au NWs elongated in the presence of a thiol solvent might, under certain conditions, change from behaving as metals to semiconductors as a result of the formation of polytetrahedral neck structures.
%We visually identify similar clusters in simulation snapshots of Au nanowires published in Figure 7 of reference \cite{Wu:2006}, calculated using the embedded atom model, and 

\section{Conclusion}
In this work, we showed good agreement between ReaxFF and DFT in terms of the absolute energy and relative energy scaling.  We also showed that ReaxFF provides a better energetic description than the semi-empirical TB-SMA potential, which has commonly been used to study Au NW elongation. We further demonstrated the formation of polytetrahedral local structures within the necks of elongated Au NWs simulated with ReaxFF.   We observed that the formation polytetrahedra is not a strong motif of structural relaxation for low temperature systems, but the likelihood of forming polytetrahedra increases with temperature.   We also found that the likelihood of forming polytetrahedra increases as NW diameter decreases.  Polytetrahedral order was also observed for simulations performed with the TB-SMA potential, however, TB-SMA appears to significantly over-predict their formation as compared to ReaxFF, especially for small diameter NWs and low temperature.  We find that ReaxFF-based simulations produce trajectories that demonstrate clear step-wise behavior in conductance that closely matches experiment. We also observe that the formation of polytetrahedra (i.e. non-crystalline structures) within the NW necks dampens the quantization behavior of the conductance, closely matching the observations in Ref. \cite{Tavazza:2011}.  The results presented here suggest that the formation of polytetrahedral structures within the necks of the [100] orientated Au NWs is a ubiquitous and important mechanism for relieving the internal stress brought about by elongation, in particular, for small diameter NWs at high temperature.  

\textit{Acknowledgment:} This work was funded by the US Department of Energy (DOE) and supported by computational resources provided by the National Energy Research Scientific Computing Center (NERSC) of the DOE under contract No. DE-AC02-05CH11231 and the National Institute for Computational Sciences, project-ID UT-TNEDU014 \cite{Vetter:2011}. WRF was supported under the U.S. Department of Education Graduate Assistance in Areas of National Need (GAANN) Fellowship under Grant No. P200A090323. Research by PRCK at the Center for Nanophase Materials Sciences was sponsored at Oak Ridge National Laboratory by the Office of Basic Energy Sciences, DOE.

\section{Methods}
\textit{DFT Calculations:} We perform DFT calculations using the generalized gradient approximation of Perdew, Burke, Ernzerhof (PBE-GGA)\cite{Perdew:1996} within the plane wave projector augmented wave formalism (PAW) \cite{Kresse:1996, Kresse:1999}. Our calculations were performed at the gamma point with a 230eV plane-wave energy cutoff. The PAW potentials included the outermost d and s electrons in the valence, giving 11 electrons per atom. Significantly, we fully included spin-orbit coupling in all our calculations of the energy and forces. All reported DFT energy values are calculated using structures that have been fully and self-consistently energy-minimized. 

\textit{Simulation Methodology:} Our basic simulation procedure closely mimics the ``stretch and relax'' procedure of Pu, \textit{et al.} \cite{Pu:2008}: we rigidly fix the last two rows of each end of the NW creating ``grip'' atoms at each end; we allow the core of the system to relax for 100 ps; after relaxation, we elongate the NW by displacing the grip atoms at one end by increments of 0.05 \AA~, allowing the system to relax for a prescribed time between displacements, repeating until the NW breaks. We explore elongation rates from 5 m/s (1 ps relaxation between displacements) to 0.1 m/s (50 ps). This range encompasses typical elongation rates used to draw metallic whiskers with STM or AFM \cite{Agrait:1995}. Simulations are conducted between T=10 K and 500 K; common experimental conditions are T=4-10 K \cite{Agrait:1995, Yanson:1998} and T$\sim$ 300K \cite{Marszalek:2000, Coura:2004}.  For each statepoint we perform 10 independent simulations to obtain sufficient statistics.

ReaxFF simulations use the fitting parameters derived by Keith, et al.\cite{Keith:2010}.  This particular fitting was chosen because it was parameterized using PBE-GGA DFT, which is well suited for systems with surfaces (i.e., NWs), and the elastic moduli of the fitting closely match experiment\cite{Keith:2010}. TB-SMA simulations are performed using fitting parameters derived by Cleri and Rosato\cite{Cleri:1993}. TB-SMA is used as previous comparisons with DFT showed it provides a better match in terms of relative energy change than other common semi-empirical potentials\cite{Pu:2007}.  Both ReaxFF and TB-SMA simulations are performed using the LAMMPS simulation package\cite{Plimpton:1995} (extended to incorporate TB-SMA) with the Nose-Hoover thermostat for systems without periodic boundary effects. As is common, TB-SMA simulations use a timestep of 2 femtoseconds \cite{Coura:2004, Sato:2005, Pu:2007, Pu:2008}, whereas a smaller timestep of 0.5 femtoseconds is needed for ReaxFF.  To simulate the same timescale, ReaxFF requires more walltime than TB-SMA, by a factor of $\sim$50.

\textit{$R_{ylm}$ method:} To analyze the local configurations of atoms, we employ techniques borrowed from the field of \textit{shape matching}\cite{Keys:2011,Keys:2011b}.  We use the R$_{ylm}$ scheme proposed in references \cite{Iacovella:2007, Iacovella:2008}, which allows us to objectively identify structural trends in an automated fashion. This method relies on creating a rotationally invariant spherical harmonics \cite{Steinhardt:1983} ``fingerprint'' of the first neighbor shell surrounding each atom. We compare the fingerprint of each local cluster to a library of reference configurations and determine the best match. Here, we quantify the match, $M$, using the Euclidian distance, normalized such that it spans from 0 (worst match) to 1 (perfect match). By definition, every cluster in this scheme has a ``best match'' even if that match is poor. To avoid misidentification, we define a cutoff such that we ignore matches where $M < 0.86$. This provides results consistent with visual inspection; the choice of cutoff will impact the specific calculated values, however the overall trends are not strongly perturbed by this choice\cite{Keys:2011}. Our reference library includes full and partially coordinated clusters of type FCC, HCP, and icoshaedra, incorporating both ideal configurations and configurations relaxed using the ReaxFF and TB-SMA potentials, where we only consider clusters of 6 $\le$ cn $\le$ 12. 

\textit{Conductance Measurement:} Conductance calculations are carried out in a combined DFT-non-equilibrium Green's function framework. A double-zeta numerical atomic orbital basis set is used (18 basis functions per Au atom) in the calculations. The zero-bias conductance is given by $G=T(E_F)$ where $T(E)$ is the transmission coefficient of the device and $E_F$ is the Fermi energy. $G$ is reported in units of $G_{0} = 2e^2/h$, where $e$ is the charge of an electron and $h$ is Planck's constant. We validated our conductance measurements by investigating single point atom contacts, finding good agreement with previous theoretical calculations \cite{Tavazza:2011, Tavazza:2010, Lopez:2008}.  Further details of the implementation and its accuracy can be found in Ref. \cite{PhysRevB.81.115412}.

%\bibliography{auwire_library.bib}

\begin{mcitethebibliography}{64}
\providecommand*{\natexlab}[1]{#1}
\providecommand*{\mciteSetBstSublistMode}[1]{}
\providecommand*{\mciteSetBstMaxWidthForm}[2]{}
\providecommand*{\mciteBstWouldAddEndPuncttrue}
  {\def\EndOfBibitem{\unskip.}}
\providecommand*{\mciteBstWouldAddEndPunctfalse}
  {\let\EndOfBibitem\relax}
\providecommand*{\mciteSetBstMidEndSepPunct}[3]{}
\providecommand*{\mciteSetBstSublistLabelBeginEnd}[3]{}
\providecommand*{\EndOfBibitem}{}
\mciteSetBstSublistMode{f}
\mciteSetBstMaxWidthForm{subitem}{(\alph{mcitesubitemcount})}
\mciteSetBstSublistLabelBeginEnd{\mcitemaxwidthsubitemform\space}
{\relax}{\relax}

\bibitem[Muller et~al.(1992)Muller, van Ruitenbeek J.~M., and de~Jongh
  L.~J.]{Muller:1992}
Muller,~C.; van Ruitenbeek J.~M.,; de~Jongh L.~J., Conductance and supercurrent
  discontinuities in atomic-scale metallic constrictions of variable width.
  \emph{Physical Review Letters} \textbf{1992}, \emph{69}, 140--143\relax
\mciteBstWouldAddEndPuncttrue
\mciteSetBstMidEndSepPunct{\mcitedefaultmidpunct}
{\mcitedefaultendpunct}{\mcitedefaultseppunct}\relax
\EndOfBibitem
\bibitem[Krans et~al.(1995)Krans, van Rutlenbeek, Fisun, Yanson, and
  de~Jongh]{Krans:1995}
Krans,~J.~M.; van Rutlenbeek,~J.~M.; Fisun,~V.~V.; Yanson,~I.~K.;
  de~Jongh,~L.~J. The signature of conductance quantization in metallic point
  contacts. \emph{Nature} \textbf{1995}, \emph{375}, 767--769\relax
\mciteBstWouldAddEndPuncttrue
\mciteSetBstMidEndSepPunct{\mcitedefaultmidpunct}
{\mcitedefaultendpunct}{\mcitedefaultseppunct}\relax
\EndOfBibitem
\bibitem[Ohnishi et~al.(1998)Ohnishi, Kondo, and Takayanagi]{Ohnishi:1998}
Ohnishi,~H.; Kondo,~Y.; Takayanagi,~K. Quantized conductance through individual
  rows of suspended gold atoms. \emph{Nature} \textbf{1998}, \emph{395},
  780--783\relax
\mciteBstWouldAddEndPuncttrue
\mciteSetBstMidEndSepPunct{\mcitedefaultmidpunct}
{\mcitedefaultendpunct}{\mcitedefaultseppunct}\relax
\EndOfBibitem
\bibitem[Yanson et~al.(1998)Yanson, Bollinger, van~den Brom, Agrait, and van
  Ruitenbeek]{Yanson:1998}
Yanson,~A.~I.; Bollinger,~G.~R.; van~den Brom,~H.~E.; Agrait,~N.; van
  Ruitenbeek,~J.~M. Formation and manipulation of a metallic wire of single
  gold atoms. \emph{Nature} \textbf{1998}, \emph{395}, 783--785\relax
\mciteBstWouldAddEndPuncttrue
\mciteSetBstMidEndSepPunct{\mcitedefaultmidpunct}
{\mcitedefaultendpunct}{\mcitedefaultseppunct}\relax
\EndOfBibitem
\bibitem[Guo et~al.(2011)Guo, Hihath, and Tao]{Guo:2011}
Guo,~S.; Hihath,~J.; Tao,~N. Breakdown of Atomic-Sized Metallic Contacts
  Measured on Nanosecond Scale. \emph{Nano Letters} \textbf{2011}, \emph{11},
  927--933\relax
\mciteBstWouldAddEndPuncttrue
\mciteSetBstMidEndSepPunct{\mcitedefaultmidpunct}
{\mcitedefaultendpunct}{\mcitedefaultseppunct}\relax
\EndOfBibitem
\bibitem[Marszalek et~al.(2000)Marszalek, Greenleaf, and Li]{Marszalek:2000}
Marszalek,~P.; Greenleaf,~W.; Li,~H. Atomic force microscopy captures quantized
  plastic deformation in gold nanowires. \emph{Proceedings of the National
  Academy of Sciences} \textbf{2000}, \emph{97}, 6282--6286\relax
\mciteBstWouldAddEndPuncttrue
\mciteSetBstMidEndSepPunct{\mcitedefaultmidpunct}
{\mcitedefaultendpunct}{\mcitedefaultseppunct}\relax
\EndOfBibitem
\bibitem[Coura et~al.(2004)Coura, Legoas, Moreira, and Sato]{Coura:2004}
Coura,~P.; Legoas,~S.; Moreira,~A.; Sato,~F. On the Structural and Stability
  Features of Linear Atomic Suspended Chains Formed from Gold Nanowires
  Stretching. \emph{Nano Letters} \textbf{2004}, \emph{4}, 1187--1191\relax
\mciteBstWouldAddEndPuncttrue
\mciteSetBstMidEndSepPunct{\mcitedefaultmidpunct}
{\mcitedefaultendpunct}{\mcitedefaultseppunct}\relax
\EndOfBibitem
\bibitem[da~Silva et~al.(2004)da~Silva, da~Silva, and Fazzio]{daSilva:2004}
da~Silva,~E.; da~Silva,~A.; Fazzio,~A. Breaking of gold nanowires.
  \emph{Computational Materials Science} \textbf{2004}, \emph{30}, 73--76\relax
\mciteBstWouldAddEndPuncttrue
\mciteSetBstMidEndSepPunct{\mcitedefaultmidpunct}
{\mcitedefaultendpunct}{\mcitedefaultseppunct}\relax
\EndOfBibitem
\bibitem[Gall et~al.(2004)Gall, Diao, and Dunn]{Gall:2004}
Gall,~K.; Diao,~J.; Dunn,~M.~L. The Strength of Gold Nanowires. \emph{Nano
  Letters} \textbf{2004}, \emph{4}, 2431--2436\relax
\mciteBstWouldAddEndPuncttrue
\mciteSetBstMidEndSepPunct{\mcitedefaultmidpunct}
{\mcitedefaultendpunct}{\mcitedefaultseppunct}\relax
\EndOfBibitem
\bibitem[Sato et~al.(2005)Sato, Moreira, Coura, Dantas, Legoas, Ugarte, and
  Galv{\~a}o]{Sato:2005}
Sato,~F.; Moreira,~A.~S.; Coura,~P.~Z.; Dantas,~S.~O.; Legoas,~S.~B.;
  Ugarte,~D.; Galv{\~a}o,~D.~S. Computer simulations of gold nanowire
  formation: the role of outlayer atoms. \emph{Applied Physics A}
  \textbf{2005}, \emph{81}, 1527--1531\relax
\mciteBstWouldAddEndPuncttrue
\mciteSetBstMidEndSepPunct{\mcitedefaultmidpunct}
{\mcitedefaultendpunct}{\mcitedefaultseppunct}\relax
\EndOfBibitem
\bibitem[Koh and Lee(2006)]{Koh:2006}
Koh,~S. J.~A.; Lee,~H.~P. Molecular dynamics simulation of size and strain rate
  dependent mechanical response of FCC metallic nanowires.
  \emph{Nanotechnology} \textbf{2006}, \emph{17}, 3451--67\relax
\mciteBstWouldAddEndPuncttrue
\mciteSetBstMidEndSepPunct{\mcitedefaultmidpunct}
{\mcitedefaultendpunct}{\mcitedefaultseppunct}\relax
\EndOfBibitem
\bibitem[Pu et~al.(2007)Pu, Leng, Tsetseris, Park, Pantelides, and
  Cummings]{Pu:2007}
Pu,~Q.; Leng,~Y.; Tsetseris,~L.; Park,~H.~S.; Pantelides,~S.~T.;
  Cummings,~P.~T. Molecular dynamics simulations of stretched gold nanowires:
  the relative utility of different semiempirical potentials. \emph{The Journal
  of Chemical Physics} \textbf{2007}, \emph{126}, 144707\relax
\mciteBstWouldAddEndPuncttrue
\mciteSetBstMidEndSepPunct{\mcitedefaultmidpunct}
{\mcitedefaultendpunct}{\mcitedefaultseppunct}\relax
\EndOfBibitem
\bibitem[Wang et~al.(2007)Wang, Zhao, Hu, Yin, Liang, Liu, and Deng]{Wang:2007}
Wang,~D.; Zhao,~J.; Hu,~S.; Yin,~X.; Liang,~S.; Liu,~Y.; Deng,~S. Where, and
  how, does a nanowire break? \emph{Nano Letters} \textbf{2007}, \emph{7},
  1208--12\relax
\mciteBstWouldAddEndPuncttrue
\mciteSetBstMidEndSepPunct{\mcitedefaultmidpunct}
{\mcitedefaultendpunct}{\mcitedefaultseppunct}\relax
\EndOfBibitem
\bibitem[Pu et~al.(2008)Pu, Leng, and Cummings]{Pu:2008}
Pu,~Q.; Leng,~Y.; Cummings,~P.~T. Rate-dependent energy release mechanism of
  gold nanowires under elongation. \emph{Journal of the American Chemical
  Society} \textbf{2008}, \emph{130}, 17907--12\relax
\mciteBstWouldAddEndPuncttrue
\mciteSetBstMidEndSepPunct{\mcitedefaultmidpunct}
{\mcitedefaultendpunct}{\mcitedefaultseppunct}\relax
\EndOfBibitem
\bibitem[Tsutsui et~al.(2008)Tsutsui, Shoji, Taniguchi, and
  Kawai]{Tsutsui:2008}
Tsutsui,~M.; Shoji,~K.; Taniguchi,~M.; Kawai,~T. Formation and self-breaking
  mechanism of stable atom-sized junctions. \emph{Nano Letters} \textbf{2008},
  \emph{8}, 345--9\relax
\mciteBstWouldAddEndPuncttrue
\mciteSetBstMidEndSepPunct{\mcitedefaultmidpunct}
{\mcitedefaultendpunct}{\mcitedefaultseppunct}\relax
\EndOfBibitem
\bibitem[Lagos et~al.(2009)Lagos, Sato, Bettini, Rodrigues, Galvao, and
  Ugarte]{Lagos:2009}
Lagos,~M.~J.; Sato,~F.; Bettini,~J.; Rodrigues,~V.; Galvao,~D.~S.; Ugarte,~D.
  Observation of the smallest metal nanotube with a square cross-section.
  \emph{Nature Nanotech.} \textbf{2009}, \emph{4}, 149--152\relax
\mciteBstWouldAddEndPuncttrue
\mciteSetBstMidEndSepPunct{\mcitedefaultmidpunct}
{\mcitedefaultendpunct}{\mcitedefaultseppunct}\relax
\EndOfBibitem
\bibitem[Tavazza et~al.(2009)Tavazza, Levine, and Chaka]{Tavazza:2009}
Tavazza,~F.; Levine,~L.; Chaka,~A. Elongation and breaking mechanisms of gold
  nanowires under a wide range of tensile conditions. \emph{Journal of Applied
  Physics} \textbf{2009}, \emph{106}, 043522\relax
\mciteBstWouldAddEndPuncttrue
\mciteSetBstMidEndSepPunct{\mcitedefaultmidpunct}
{\mcitedefaultendpunct}{\mcitedefaultseppunct}\relax
\EndOfBibitem
\bibitem[Barnett and Landman(1997)]{Barnett:1997}
Barnett,~R.~N.; Landman,~U. Cluster-derived structures and conductance
  fluctuations in nanowires. \emph{Nature} \textbf{1997}, \emph{387},
  788--791\relax
\mciteBstWouldAddEndPuncttrue
\mciteSetBstMidEndSepPunct{\mcitedefaultmidpunct}
{\mcitedefaultendpunct}{\mcitedefaultseppunct}\relax
\EndOfBibitem
\bibitem[Rego et~al.(2003)Rego, Rocha, Rodrigues, and Ugarte]{Rego:2003}
Rego,~L. G.~C.; Rocha,~A.~R.; Rodrigues,~V.; Ugarte,~D. Role of structural
  evolution in the quantum conductance behavior of gold nanowires during
  stretching. \emph{Physical Review B} \textbf{2003}, \emph{67}, 045412\relax
\mciteBstWouldAddEndPuncttrue
\mciteSetBstMidEndSepPunct{\mcitedefaultmidpunct}
{\mcitedefaultendpunct}{\mcitedefaultseppunct}\relax
\EndOfBibitem
\bibitem[Lopez-Acevedo et~al.(2008)Lopez-Acevedo, Koudela, and
  H{\"a}kkinen]{Lopez:2008}
Lopez-Acevedo,~O.; Koudela,~D.; H{\"a}kkinen,~H. Conductance through atomic
  point contacts between fcc(100) electrodes of gold. \emph{The European
  Physical Journal B} \textbf{2008}, \emph{66}, 497--501\relax
\mciteBstWouldAddEndPuncttrue
\mciteSetBstMidEndSepPunct{\mcitedefaultmidpunct}
{\mcitedefaultendpunct}{\mcitedefaultseppunct}\relax
\EndOfBibitem
\bibitem[Tavazza et~al.(2011)Tavazza, Smith, Levine, Pratt, and
  Chaka]{Tavazza:2011}
Tavazza,~F.; Smith,~D.~T.; Levine,~L.~E.; Pratt,~J.~R.; Chaka,~A.~M. Electron
  Transport in Gold Nanowires: Stable 1-, 2- and 3-Dimensional Atomic
  Structures and Noninteger Conduction States. \emph{Phys. Rev. Lett.}
  \textbf{2011}, \emph{107}, 126802\relax
\mciteBstWouldAddEndPuncttrue
\mciteSetBstMidEndSepPunct{\mcitedefaultmidpunct}
{\mcitedefaultendpunct}{\mcitedefaultseppunct}\relax
\EndOfBibitem
\bibitem[French et~al.(2011)French, Iacovella, and Cummings]{French:2011}
French,~W.~R.; Iacovella,~C.~R.; Cummings,~P.~T. The Influence of Molecular
  Adsorption on Elongating Gold Nanowires. \emph{The Journal of Physical
  Chemistry C} \textbf{2011}, \emph{115}, 18422--18433\relax
\mciteBstWouldAddEndPuncttrue
\mciteSetBstMidEndSepPunct{\mcitedefaultmidpunct}
{\mcitedefaultendpunct}{\mcitedefaultseppunct}\relax
\EndOfBibitem
\bibitem[Fioravante and Nunes(2007)]{Nunes:2007}
Fioravante,~F.; Nunes,~R.~W. Semiconducting chains of gold and silver.
  \emph{Applied Physics Letters} \textbf{2007}, \emph{91}, 223115\relax
\mciteBstWouldAddEndPuncttrue
\mciteSetBstMidEndSepPunct{\mcitedefaultmidpunct}
{\mcitedefaultendpunct}{\mcitedefaultseppunct}\relax
\EndOfBibitem
\bibitem[Park et~al.(2009)Park, Cai, Espinosa, and Huang]{Park:2009}
Park,~H.~S.; Cai,~W.; Espinosa,~H.~D.; Huang,~H. Mechanics of Crystalline
  Nanowires. \emph{MRS BULLETIN} \textbf{2009}, \emph{34}, 178--183\relax
\mciteBstWouldAddEndPuncttrue
\mciteSetBstMidEndSepPunct{\mcitedefaultmidpunct}
{\mcitedefaultendpunct}{\mcitedefaultseppunct}\relax
\EndOfBibitem
\bibitem[Ikeda et~al.(1999)Ikeda, Qi, \ifmmode~\mbox{\c{C}}\else
  \c{C}\fi{}agin, Samwer, Johnson, and Goddard]{Ikeda:1999}
Ikeda,~H.; Qi,~Y.; \ifmmode~\mbox{\c{C}}\else \c{C}\fi{}agin,~T.; Samwer,~K.;
  Johnson,~W.~L.; Goddard,~W.~A. Strain Rate Induced Amorphization in Metallic
  Nanowires. \emph{Physical Review Letters} \textbf{1999}, \emph{82},
  2900--2903\relax
\mciteBstWouldAddEndPuncttrue
\mciteSetBstMidEndSepPunct{\mcitedefaultmidpunct}
{\mcitedefaultendpunct}{\mcitedefaultseppunct}\relax
\EndOfBibitem
\bibitem[Zhang et~al.(2009)Zhang, Srolovitz, Douglas, and Warren]{Zhang:2009}
Zhang,~H.; Srolovitz,~D.~J.; Douglas,~J.~F.; Warren,~J.~A. Grain boundaries
  exhibit the dynamics of glass-forming liquids. \emph{Proceedings of the
  National Academy of Sciences} \textbf{2009}, \emph{106}, 7735--7740\relax
\mciteBstWouldAddEndPuncttrue
\mciteSetBstMidEndSepPunct{\mcitedefaultmidpunct}
{\mcitedefaultendpunct}{\mcitedefaultseppunct}\relax
\EndOfBibitem
\bibitem[Vel{\'a}zquez-Salazar et~al.(2011)Vel{\'a}zquez-Salazar, Esparza,
  Mej{\'\i}a-Rosales, Estrada-Salas, Ponce, Deepak, Castro-Guerrero, and
  Jos{\'e}-Yacam{\'a}n]{Jesus:2011}
Vel{\'a}zquez-Salazar,~J.~J.; Esparza,~R.; Mej{\'\i}a-Rosales,~S.~J.;
  Estrada-Salas,~R.; Ponce,~A.; Deepak,~F.~L.; Castro-Guerrero,~C.;
  Jos{\'e}-Yacam{\'a}n,~M. Experimental Evidence of Icosahedral and Decahedral
  Packing in One-Dimensional Nanostructures. \emph{ACS Nano} \textbf{2011},
  \emph{5}, 6272--6278\relax
\mciteBstWouldAddEndPuncttrue
\mciteSetBstMidEndSepPunct{\mcitedefaultmidpunct}
{\mcitedefaultendpunct}{\mcitedefaultseppunct}\relax
\EndOfBibitem
\bibitem[Jiang et~al.(2009)Jiang, Nobusada, Luo, and Whetten]{Jiang:2009}
Jiang,~D.-e.; Nobusada,~K.; Luo,~W.; Whetten,~R.~L. Thiolated Gold Nanowires:
  Metallic versus Semiconducting. \emph{ACS Nano} \textbf{2009}, \emph{3},
  2351--2357, PMID: 19603760\relax
\mciteBstWouldAddEndPuncttrue
\mciteSetBstMidEndSepPunct{\mcitedefaultmidpunct}
{\mcitedefaultendpunct}{\mcitedefaultseppunct}\relax
\EndOfBibitem
\bibitem[Iacovella et~al.(2007)Iacovella, Keys, Horsch, and
  Glotzer]{Iacovella:2007}
Iacovella,~C.~R.; Keys,~A.~S.; Horsch,~M.~A.; Glotzer,~S.~C. Icosahedral
  packing of polymer-tethered nanospheres and stabilization of the gyroid
  phase. \emph{Physical Review E} \textbf{2007}, \emph{75}, 040801\relax
\mciteBstWouldAddEndPuncttrue
\mciteSetBstMidEndSepPunct{\mcitedefaultmidpunct}
{\mcitedefaultendpunct}{\mcitedefaultseppunct}\relax
\EndOfBibitem
\bibitem[Keys et~al.(2011)Keys, Iacovella, and Glotzer]{Keys:2011}
Keys,~A.~S.; Iacovella,~C.~R.; Glotzer,~S.~C. Characterizing Structure Through
  Shape Matching and Applications to Self-Assembly. \emph{Annual Review of
  Condensed Matter Physics} \textbf{2011}, \emph{2}, 263--285\relax
\mciteBstWouldAddEndPuncttrue
\mciteSetBstMidEndSepPunct{\mcitedefaultmidpunct}
{\mcitedefaultendpunct}{\mcitedefaultseppunct}\relax
\EndOfBibitem
\bibitem[Keys et~al.(2011)Keys, Iacovella, and Glotzer]{Keys:2011b}
Keys,~A.~S.; Iacovella,~C.~R.; Glotzer,~S.~C. Characterizing complex particle
  morphologies through shape matching: Descriptors, applications, and
  algorithms. \emph{Journal of Computational Physics} \textbf{2011},
  \emph{230}, 6438 -- 6463\relax
\mciteBstWouldAddEndPuncttrue
\mciteSetBstMidEndSepPunct{\mcitedefaultmidpunct}
{\mcitedefaultendpunct}{\mcitedefaultseppunct}\relax
\EndOfBibitem
\bibitem[H{\"a}kkinen et~al.(2000)H{\"a}kkinen, Barnett, Scherbakov, and
  Landman]{Hakkinen:2000}
H{\"a}kkinen,~H.; Barnett,~R.~N.; Scherbakov,~A.~G.; Landman,~U. Nanowire Gold
  Chains:  Formation Mechanisms and Conductance. \emph{The Journal of
  Physical Chemistry B} \textbf{2000}, \emph{104}, 9063--9066\relax
\mciteBstWouldAddEndPuncttrue
\mciteSetBstMidEndSepPunct{\mcitedefaultmidpunct}
{\mcitedefaultendpunct}{\mcitedefaultseppunct}\relax
\EndOfBibitem
\bibitem[V{\'e}lez et~al.(2008)V{\'e}lez, Dassie, and Leiva]{Velez:2008}
V{\'e}lez,~P.; Dassie,~S.~A.; Leiva,~E. P.~M. When do nanowires break? A model
  for the theoretical study of the long-term stability of monoatomic nanowires.
  \emph{Chemical Physics Letters} \textbf{2008}, \emph{460}, 261--265\relax
\mciteBstWouldAddEndPuncttrue
\mciteSetBstMidEndSepPunct{\mcitedefaultmidpunct}
{\mcitedefaultendpunct}{\mcitedefaultseppunct}\relax
\EndOfBibitem
\bibitem[Tavazza et~al.(2010)Tavazza, Levine, and Chaka]{Tavazza:2010}
Tavazza,~F.; Levine,~L.~E.; Chaka,~A.~M. Structural changes during the
  formation of gold single-atom chains: Stability criteria and electronic
  structure. \emph{Phys. Rev. B} \textbf{2010}, \emph{81}, 1--12\relax
\mciteBstWouldAddEndPuncttrue
\mciteSetBstMidEndSepPunct{\mcitedefaultmidpunct}
{\mcitedefaultendpunct}{\mcitedefaultseppunct}\relax
\EndOfBibitem
\bibitem[Wu(2006)]{Wu:2006}
Wu,~H. Molecular dynamics study of the mechanics of metal nanowires at finite
  temperature. \emph{European Journal of Mechanics-A/Solids} \textbf{2006},
  370--377\relax
\mciteBstWouldAddEndPuncttrue
\mciteSetBstMidEndSepPunct{\mcitedefaultmidpunct}
{\mcitedefaultendpunct}{\mcitedefaultseppunct}\relax
\EndOfBibitem
\bibitem[Zhao et~al.(2008)Zhao, Murakoshi, Yin, Kiguchi, and
  {\ldots}]{Zhao:2008}
Zhao,~J.; Murakoshi,~K.; Yin,~X.; Kiguchi,~M.; {\ldots},~Y.~G. Dynamic
  Characterization of the Postbreaking Behavior of a Nanowire. \emph{The
  Journal of Physical Chemistry C} \textbf{2008}, \emph{112},
  20088--20094\relax
\mciteBstWouldAddEndPuncttrue
\mciteSetBstMidEndSepPunct{\mcitedefaultmidpunct}
{\mcitedefaultendpunct}{\mcitedefaultseppunct}\relax
\EndOfBibitem
\bibitem[Xiao and Wang(2004)]{Xiao:2004}
Xiao,~L.; Wang,~L. From planar to three-dimensional structural transition in
  gold clusters and the spin-orbit coupling effect. \emph{Chemical Physics
  Letters} \textbf{2004}, \emph{392}, 452--455\relax
\mciteBstWouldAddEndPuncttrue
\mciteSetBstMidEndSepPunct{\mcitedefaultmidpunct}
{\mcitedefaultendpunct}{\mcitedefaultseppunct}\relax
\EndOfBibitem
\bibitem[Keith et~al.(2010)Keith, Fantauzzi, Jacob, and van Duin]{Keith:2010}
Keith,~J.~A.; Fantauzzi,~D.; Jacob,~T.; van Duin,~A. C.~T. Reactive forcefield
  for simulating gold surfaces and nanoparticles. \emph{Phys. Rev. B}
  \textbf{2010}, \emph{81}, 235404\relax
\mciteBstWouldAddEndPuncttrue
\mciteSetBstMidEndSepPunct{\mcitedefaultmidpunct}
{\mcitedefaultendpunct}{\mcitedefaultseppunct}\relax
\EndOfBibitem
\bibitem[Cleri and Rosato(1993)]{Cleri:1993}
Cleri,~F.; Rosato,~V. Tight-binding potentials for transition metals and
  alloys. \emph{Physical Review B} \textbf{1993}, \emph{48}, 22--33\relax
\mciteBstWouldAddEndPuncttrue
\mciteSetBstMidEndSepPunct{\mcitedefaultmidpunct}
{\mcitedefaultendpunct}{\mcitedefaultseppunct}\relax
\EndOfBibitem
\bibitem[J{\"a}rvi et~al.(2008)J{\"a}rvi, Kuronen, Hakala, Nordlund, van Duin,
  Goddard, and Jacob]{Jarvi:2008}
J{\"a}rvi,~T.~T.; Kuronen,~A.; Hakala,~M.; Nordlund,~K.; van Duin,~A.~C.;
  Goddard,~W.~A.; Jacob,~T. Development of a ReaxFF description for gold.
  \emph{The European Physical Journal B - Condensed Matter and Complex Systems}
  \textbf{2008}, \emph{66}, 75--79, 10.1140/epjb/e2008-00378-3\relax
\mciteBstWouldAddEndPuncttrue
\mciteSetBstMidEndSepPunct{\mcitedefaultmidpunct}
{\mcitedefaultendpunct}{\mcitedefaultseppunct}\relax
\EndOfBibitem
\bibitem[Gall et~al.(2005)Gall, Diao, Dunn, Haftel, Bernstein, and
  Mehl]{Gall:2005}
Gall,~K.; Diao,~J.; Dunn,~M.~L.; Haftel,~M.; Bernstein,~N.; Mehl,~M.~J.
  Tetragonal Phase Transformation in Gold Nanowires. \emph{Journal of
  Engineering Materials and Technology} \textbf{2005}, \emph{127},
  417--422\relax
\mciteBstWouldAddEndPuncttrue
\mciteSetBstMidEndSepPunct{\mcitedefaultmidpunct}
{\mcitedefaultendpunct}{\mcitedefaultseppunct}\relax
\EndOfBibitem
\bibitem[van Duin et~al.(2001)van Duin, Dasgupta, Lorant, and Goddard]{goddard}
van Duin,~A. C.~T.; Dasgupta,~S.; Lorant,~F.; Goddard,~W.~A. ReaxFF:  A
  Reactive Force Field for Hydrocarbons. \emph{The Journal of Physical
  Chemistry A} \textbf{2001}, \emph{105}, 9396--9409\relax
\mciteBstWouldAddEndPuncttrue
\mciteSetBstMidEndSepPunct{\mcitedefaultmidpunct}
{\mcitedefaultendpunct}{\mcitedefaultseppunct}\relax
\EndOfBibitem
\bibitem[Plimpton(1995)]{Plimpton:1995}
Plimpton,~S.~J. Fast Parallel Algorithms for Short-Range Molecular Dynamic.
  \emph{Journal of Computational Physics} \textbf{1995}, \emph{117},
  1--19\relax
\mciteBstWouldAddEndPuncttrue
\mciteSetBstMidEndSepPunct{\mcitedefaultmidpunct}
{\mcitedefaultendpunct}{\mcitedefaultseppunct}\relax
\EndOfBibitem
\bibitem[Frank and Kasper(1958)]{Frank:1958}
Frank,~F.; Kasper,~J. Complex alloy structures regarded as sphere packings. I.
  Definitions and basic principles. \emph{Acta Crystallographica}
  \textbf{1958}, \emph{11}, 184--190\relax
\mciteBstWouldAddEndPuncttrue
\mciteSetBstMidEndSepPunct{\mcitedefaultmidpunct}
{\mcitedefaultendpunct}{\mcitedefaultseppunct}\relax
\EndOfBibitem
\bibitem[Iacovella et~al.(2008)Iacovella, Horsch, and Glotzer]{Iacovella:2008}
Iacovella,~C.~R.; Horsch,~M.~A.; Glotzer,~S.~C. Local ordering of
  polymer-tethered nanospheres and nanorods and the stabilization of the double
  gyroid phase. \emph{The Journal of Chemical Physics} \textbf{2008},
  \emph{129}, 044902\relax
\mciteBstWouldAddEndPuncttrue
\mciteSetBstMidEndSepPunct{\mcitedefaultmidpunct}
{\mcitedefaultendpunct}{\mcitedefaultseppunct}\relax
\EndOfBibitem
\bibitem[Steinhardt et~al.(1983)Steinhardt, Nelson, and
  Ronchetti]{Steinhardt:1983}
Steinhardt,~P.~J.; Nelson,~D.~R.; Ronchetti,~M. Bond-Orientational Order in
  Liquids and Glasses. \emph{Physical Review B} \textbf{1983}, \emph{28},
  784--805\relax
\mciteBstWouldAddEndPuncttrue
\mciteSetBstMidEndSepPunct{\mcitedefaultmidpunct}
{\mcitedefaultendpunct}{\mcitedefaultseppunct}\relax
\EndOfBibitem
\bibitem[Wang et~al.(2001)Wang, Yin, Wang, Buldum, and Zhao]{Wang:2001}
Wang,~B.; Yin,~S.; Wang,~G.; Buldum,~A.; Zhao,~J. Novel structures and
  properties of gold nanowires. \emph{Physical Review Letters} \textbf{2001},
  \emph{86}, 2046--2049\relax
\mciteBstWouldAddEndPuncttrue
\mciteSetBstMidEndSepPunct{\mcitedefaultmidpunct}
{\mcitedefaultendpunct}{\mcitedefaultseppunct}\relax
\EndOfBibitem
\bibitem[Driscoll and Varga(2010)]{PhysRevB.81.115412}
Driscoll,~J.~A.; Varga,~K. Convergence in quantum transport calculations:
  Localized atomic orbitals versus nonlocalized basis sets. \emph{Phys. Rev. B}
  \textbf{2010}, \emph{81}, 115412\relax
\mciteBstWouldAddEndPuncttrue
\mciteSetBstMidEndSepPunct{\mcitedefaultmidpunct}
{\mcitedefaultendpunct}{\mcitedefaultseppunct}\relax
\EndOfBibitem
\bibitem[Costa-Kr\"amer(1997)]{costakramer:1997}
Costa-Kr\"amer,~J.~L. Conductance quantization at room temperature in magnetic
  and nonmagnetic metallic nanowires. \emph{Phys. Rev. B} \textbf{1997},
  \emph{55}, R4875--R4878\relax
\mciteBstWouldAddEndPuncttrue
\mciteSetBstMidEndSepPunct{\mcitedefaultmidpunct}
{\mcitedefaultendpunct}{\mcitedefaultseppunct}\relax
\EndOfBibitem
\bibitem[Li et~al.(2007)Li, Wang, Yang, Zhu, and Tang]{Li:2007}
Li,~X.-B.; Wang,~H.-Y.; Yang,~X.-D.; Zhu,~Z.-H.; Tang,~Y.-J. Size dependence of
  the structures and energetic and electronic properties of gold clusters.
  \emph{Journal of Chemical Physics} \textbf{2007}, \emph{126}, 084505\relax
\mciteBstWouldAddEndPuncttrue
\mciteSetBstMidEndSepPunct{\mcitedefaultmidpunct}
{\mcitedefaultendpunct}{\mcitedefaultseppunct}\relax
\EndOfBibitem
\bibitem[Gruber et~al.(2008)Gruber, Heimel, Romaner, Br\'edas, and
  Zojer]{Gruber:2008}
Gruber,~M.; Heimel,~G.; Romaner,~L.; Br\'edas,~J.-L.; Zojer,~E.
  First-principles study of the geometric and electronic structure of
  ${\mathrm{Au}}_{13}$ clusters: Importance of the prism motif. \emph{Phys.
  Rev. B} \textbf{2008}, \emph{77}, 165411\relax
\mciteBstWouldAddEndPuncttrue
\mciteSetBstMidEndSepPunct{\mcitedefaultmidpunct}
{\mcitedefaultendpunct}{\mcitedefaultseppunct}\relax
\EndOfBibitem
\bibitem[Pundlik et~al.(2011)Pundlik, Kalyanaraman, and Waghmare]{Pundlik:2011}
Pundlik,~S.~S.; Kalyanaraman,~K.; Waghmare,~U.~V. First-Principles
  Investigation of the Atomic and Electronic Structure and Magnetic Moments in
  Gold Nanoclusters. \emph{The Journal of Physical Chemistry C} \textbf{2011},
  \emph{115}, 3809--3820\relax
\mciteBstWouldAddEndPuncttrue
\mciteSetBstMidEndSepPunct{\mcitedefaultmidpunct}
{\mcitedefaultendpunct}{\mcitedefaultseppunct}\relax
\EndOfBibitem
\bibitem[Menard et~al.(2006)Menard, Xu, Gao, Twesten, Harper, Song, Wang,
  Douglas, Yang, Frenkel, Murray, and Nuzzo]{Menard:2006}
Menard,~L.~D.; Xu,~H.; Gao,~S.-P.; Twesten,~R.~D.; Harper,~A.~S.; Song,~Y.;
  Wang,~G.; Douglas,~A.~D.; Yang,~J.~C.; Frenkel,~A.~I. et~al. Metal Core
  Bonding Motifs of Monodisperse Icosahedral Au13 and Larger Au
  Monolayer-Protected Clusters As Revealed by X-ray Absorption Spectroscopy and
  Transmission Electron Microscopy. \emph{The Journal of Physical Chemistry B}
  \textbf{2006}, \emph{110}, 14564--14573\relax
\mciteBstWouldAddEndPuncttrue
\mciteSetBstMidEndSepPunct{\mcitedefaultmidpunct}
{\mcitedefaultendpunct}{\mcitedefaultseppunct}\relax
\EndOfBibitem
\bibitem[Walter et~al.(2008)Walter, Akola, and Lopez-Acevedo]{Walter:2008}
Walter,~M.; Akola,~J.; Lopez-Acevedo,~O. A unified view of ligand-protected
  gold clusters as superatom complexes. \emph{Proceedings of the National
  Academy of Sciences} \textbf{2008}, \emph{105}, 9157--9162\relax
\mciteBstWouldAddEndPuncttrue
\mciteSetBstMidEndSepPunct{\mcitedefaultmidpunct}
{\mcitedefaultendpunct}{\mcitedefaultseppunct}\relax
\EndOfBibitem
\bibitem[Pei et~al.(2008)Pei, Gao, and Zeng]{Pei:2008}
Pei,~Y.; Gao,~Y.; Zeng,~X.~C. Structural Prediction of Thiolate-Protected Au38:
  A Face-Fused Bi-icosahedral Au Core. \emph{Journal of the American Chemical
  Society} \textbf{2008}, \emph{130}, 7830--7832\relax
\mciteBstWouldAddEndPuncttrue
\mciteSetBstMidEndSepPunct{\mcitedefaultmidpunct}
{\mcitedefaultendpunct}{\mcitedefaultseppunct}\relax
\EndOfBibitem
\bibitem[Shichibu et~al.(2007)Shichibu, Negishi, Watanabe, Chaki, Kawaguchi,
  and Tsukuda]{Shichibu:2007}
Shichibu,~Y.; Negishi,~Y.; Watanabe,~T.; Chaki,~N.~K.; Kawaguchi,~H.;
  Tsukuda,~T. Biicosahedral Gold Clusters [Au25(PPh3)10(SCnH2n+1)5Cl2]2+ (n =
  2−18):  A Stepping Stone to Cluster-Assembled Materials. \emph{The
  Journal of Physical Chemistry C} \textbf{2007}, \emph{111}, 7845--7847\relax
\mciteBstWouldAddEndPuncttrue
\mciteSetBstMidEndSepPunct{\mcitedefaultmidpunct}
{\mcitedefaultendpunct}{\mcitedefaultseppunct}\relax
\EndOfBibitem
\bibitem[Agrait et~al.(1995)Agrait, Rubio, and Vieira]{Agrait:1995}
Agrait,~N.; Rubio,~G.; Vieira,~S. Plastic deformation of nanometer-scale gold
  connective necks. \emph{Physical Review Letters} \textbf{1995}, \emph{74},
  3995--3998\relax
\mciteBstWouldAddEndPuncttrue
\mciteSetBstMidEndSepPunct{\mcitedefaultmidpunct}
{\mcitedefaultendpunct}{\mcitedefaultseppunct}\relax
\EndOfBibitem
\bibitem[Zheng et~al.(1990)Zheng, Hoffmann, and Nelson]{Zheng:1990}
Zheng,~C.; Hoffmann,~R.; Nelson,~D.~R. A helical face-sharing tetrahedron chain
  from irrational twist, stella quadrangula, and related matters. \emph{J Am
  Chem Soc} \textbf{1990}, \emph{112}, 3784--3791\relax
\mciteBstWouldAddEndPuncttrue
\mciteSetBstMidEndSepPunct{\mcitedefaultmidpunct}
{\mcitedefaultendpunct}{\mcitedefaultseppunct}\relax
\EndOfBibitem
\bibitem[Lidin and Andersson(1996)]{Lidin:1996}
Lidin,~S.; Andersson,~S. Regular polyhedra helices. \emph{ZEITSCHRIFT FUR
  ANORGANISCHE UND ALLGEMEINE CHEMIE} \textbf{1996}, \emph{622}, 164--166\relax
\mciteBstWouldAddEndPuncttrue
\mciteSetBstMidEndSepPunct{\mcitedefaultmidpunct}
{\mcitedefaultendpunct}{\mcitedefaultseppunct}\relax
\EndOfBibitem
\bibitem[Vetter et~al.(2011)Vetter, Glassbrook, Dongarra, Schwan, Loftis,
  McNally, Meredith, Rogers, Roth, Spafford, and Yalamanchili]{Vetter:2011}
Vetter,~J.; Glassbrook,~R.; Dongarra,~J.; Schwan,~K.; Loftis,~B.; McNally,~S.;
  Meredith,~J.; Rogers,~J.; Roth,~P.; Spafford,~K. et~al. Keeneland: Bringing
  heterogeneous GPU computing to the computational science community.
  \emph{IEEE Computing in Science and Engineering} \textbf{2011}, \emph{13},
  90--95\relax
\mciteBstWouldAddEndPuncttrue
\mciteSetBstMidEndSepPunct{\mcitedefaultmidpunct}
{\mcitedefaultendpunct}{\mcitedefaultseppunct}\relax
\EndOfBibitem
\bibitem[Perdew et~al.(1996)Perdew, Burke, and Ernzerhof]{Perdew:1996}
Perdew,~J.~P.; Burke,~K.; Ernzerhof,~M. Generalized Gradient Approximation Made
  Simple. \emph{Phys. Rev. Lett.} \textbf{1996}, \emph{77}, 3865\relax
\mciteBstWouldAddEndPuncttrue
\mciteSetBstMidEndSepPunct{\mcitedefaultmidpunct}
{\mcitedefaultendpunct}{\mcitedefaultseppunct}\relax
\EndOfBibitem
\bibitem[Kresse and Furthmuller(1996)]{Kresse:1996}
Kresse,~G.; Furthmuller,~J. Efficient iterative schemes for ab initio
  total-energy calculations using a plane-wave basis set. \emph{Phys. Rev. B}
  \textbf{1996}, \emph{54}, 11169\relax
\mciteBstWouldAddEndPuncttrue
\mciteSetBstMidEndSepPunct{\mcitedefaultmidpunct}
{\mcitedefaultendpunct}{\mcitedefaultseppunct}\relax
\EndOfBibitem
\bibitem[Kresse and Joubert(1999)]{Kresse:1999}
Kresse,~G.; Joubert,~D. From ultrasoft pseudopotentials to the projector
  augmented-wave method. \emph{Phys. Rev. B} \textbf{1999}, \emph{59},
  1758\relax
\mciteBstWouldAddEndPuncttrue
\mciteSetBstMidEndSepPunct{\mcitedefaultmidpunct}
{\mcitedefaultendpunct}{\mcitedefaultseppunct}\relax
\EndOfBibitem
\end{mcitethebibliography}
\providecommand*{\mcitethebibliography}{\thebibliography}
\csname @ifundefined\endcsname{endmcitethebibliography}
{\let\endmcitethebibliography\endthebibliography}{}

\begin{tocentry}
\includegraphics{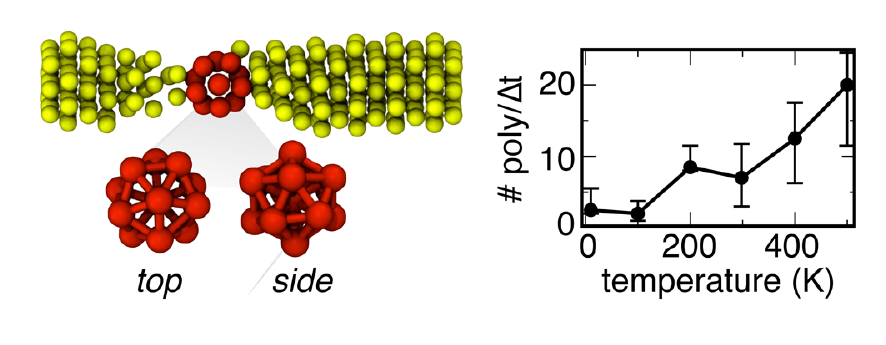}
\\\\Polytetrahedral structures form as Au nanowires undergo elongation.
\end{tocentry}

\end{document}